\documentclass{appolb}
\usepackage{amsmath}
\usepackage{graphicx}
\usepackage{color}

\begin{document}
\title{Neutron star structure and nuclear matter properties from a general Walecka-type model with Bayesian analysis
\thanks{Presented at Excited QCD 2026, held in Granada, Spain}%
}
\author{Yao Ma
\address{School of Frontier Sciences, Nanjing University, Suzhou 215163, China}
\\[3mm]
{Jia-Ying Xiong\thanks{E-mail: xiongjiaying21@mails.ucas.ac.cn}
 \address{School of Frontier Sciences, Nanjing University, Suzhou 215163, China}
 \address{School of Fundamental Physics and Mathematical Sciences, Hangzhou Institute for Advanced Study, UCAS, Hangzhou 310024, China}
}
}
\maketitle
\begin{abstract}
We establish a Bayesian analysis framework with a general Walecka-type relativistic mean-field model to study dense nuclear matter under constraints from nuclear matter properties and neutron star observations.
With experimental and observational data well described, we find that pure hadronic descriptions can generate a peak structure in sound velocity by \(\omega\), \(\rho\), \(\sigma\), and \(a_0\) meson mixing, which is crucial for describing both medium and massive neutron stars.
As the peak structure is frequently interpreted as a signature of phase transitions, our findings provide a new perspective on the microscopic origin of the sound velocity peak just with pure hadronic matter.
\end{abstract}
  
\section{Introduction}
The field of nuclear physics has advanced significantly with the discovery of GW170817~\cite{LIGOScientific:2017vwq} from a binary neutron star inspiral.
Combining with electromagnetic observations~\cite{Hessels:2004ps,Lynch:2012vv,Fonseca:2021wxt}, one can constrain neutron star (NS) structures such as mass-radius (M-R) relations~\cite{LIGOScientific:2017vwq} and tidal deformation. NS structures relate to the equation of state (EOS) of nuclear matter (NM)~\cite{Douchin:2001sv,Read:2008iy}, allowing us to probe nuclear matter at extreme regimes. Nuclear physics experiments, such as heavy-ion collisions and nuclei structure analysis~\cite{Youngblood:1999zza,Brown:2000pd}, provide constraints on NM properties around saturation density.

Theoretical developments, such as Walecka-type relativistic mean-field (RMF) models~\cite{Walecka:1974qa,Serot:1984ey}, Skyrme-Hartree-Fock models~\cite{Vautherin:1971aw}, and chiral effective field theory~\cite{Epelbaum:2008ga}, have made significant progress in describing NM properties. However, combining constraints from different sources remains challenging due to complex interplay of nuclear forces and non-perturbative QCD.

We present an AI-driven Bayesian analysis platform to search for optimal parameter space of a general Walecka-type RMF model~\cite{Guo:2024nzi}. This work develops a generalized Bayesian analysis scheme incorporating all mesons lighter than $1~\rm GeV$ with constraints from nuclei structures, heavy-ion collisions, and astrophysical observations.

We find that pure hadronic descriptions can generate a peak structure in sound velocity by $\omega$, $\rho$, $\sigma$, and $a_0$ meson mixing, which is crucial for describing both medium and massive neutron stars.
As a result, the optimal parameter space can simultaneously describe the NM properties around saturation density and NS structures.

The proceeding is organized as follows.
In Sec. II, we introduce the Bayesian analysis framework and experimental constraints.
In Sec. III, we present a general Walecka-type model for dense nuclear matter and its numerical results.
Finally, we summarize our work in the last section.

\section{The analysis framework}
In this section, the basics of Bayesian analysis are first introduced, then we describe how to incorporate the constraints from NM properties and NS observations into the Bayesian formulation with a general Walecka RMF model.

\subsection{The Bayesian analysis framework and experimental constraints}

The Bayesian analysis can provide a systematic analysis framework to incorporate various constraints from nuclear physics experiments and astrophysical observations based on Bayes' theorem~\cite{bayes1763essay},
\begin{equation}
    p(\boldsymbol{\theta} \mid \mathbf{D})=\frac{p(\mathbf{D} \mid \boldsymbol{\theta}) p(\boldsymbol{\theta})}{p(\mathbf{D})}\ ,
\end{equation}
where \(\boldsymbol{\theta}\) denotes the model parameters, \(p(\boldsymbol{\theta})\) is the prior distribution encoding theoretical expectations and empirical knowledge about the parameters before considering any data, \(p(\mathbf{D} \mid \boldsymbol{\theta})\) is the likelihood function quantifying the compatibility between the model predictions and the data, and \(p(\mathbf{D})\) is the Bayesian evidence serving as a normalization constant, defined as 
\begin{equation}
    \label{eq:evidence}
    p(\mathbf{D})=\int p(\mathbf{D} \mid \boldsymbol{\theta}) p(\boldsymbol{\theta}) d \boldsymbol{\theta}\ .
\end{equation}
With the assumption that the different data are conditionally independent, the overall likelihood can be written as the product of the likelihood for each individual data:
\begin{equation}
    \label{eq:likelihood}
    p(\mathbf{D} \mid \boldsymbol{\theta})=\prod_{i=1}^n p\left(D_i \mid \boldsymbol{\theta}\right)\ .
\end{equation}

In this work, the data \(\mathbf{D}\) consist of two parts: the NM properties and the M-R relations of neutron stars.
More specifically, the following NM properties around saturation density \(n_0\) are considered, including binding energy \(e(n)\), pressure (\(P(n)\)), incompressibility (\(K(n)\)), symmetry energy (\(E_{\rm sym}(n)\)), and slope of symmetry energy \(L(n)\), and definitions can be found in Ref.~\cite{Ma:2025llw}.
Since these NM properties are derived from different theoretical models and experiments~\cite{Sedrakian:2022ata, Dutra:2012mb, Dutra:2014qga, Cozma:2017bre, Youngblood:1999zza, Howard:2019qnv, Li:2022suc, Stone:2014wza, Pattnaik:2025gtj, Shlomo:2006ole, Garg:2018uam}, there are two main sources of net uncertainties: I. theoretical model choices, and II. experimental setup dependence.
So, in the following discussion, we summarize the constraints on NM properties as reasonable ranges supported by current theoretical and experimental knowledge, and assign it as a normal distribution with one standard deviation for simplicity.
Then, for the constraints of M-R relations of NSs, the Refs.~\cite{LIGOScientific:2018cki,Antoniadis:2013pzd,Salmi:2024aum,Vinciguerra:2023qxq,Choudhury:2024xbk,Mauviard:2025dmd,LIGOScientific:2017vwq} provide the regions for the normal distributions with corresponding one standard deviations.
It should be noted that the contribution of M-R relation to the Eq.~\eqref{eq:likelihood} is estimated by choosing the maximum value of a M-R line predicted by the model for a given constraints in the analysis of this work.

\subsection{A general Walecka-type model for dense nuclear matter}
The theoretical model used in this work to describe dense nuclear matter is a genenel Walecka-type relativistic mean-field (RMF) model that incorporates the relevant degrees of freedom below \(1\mathrm{GeV}\), including the \(\sigma\), \(\omega\), \(\rho\), and \(a_0\) mesons, as well as the nucleons (\(n\) and \(p\)), where \(\pi\) meson is neglected in the RMF approximation.
The details of the model can be found in Ref.~\cite{Ma:2026kun}.
The EOS and M-R relation can be obtained by calculating the corresponding RMF Hamiltonian and Tolman-Oppenheimer-Volkoff (TOV) equation~\cite{Oppenheimer:1939ne,Tolman:1939jz}, and computation methods can be found in Ref.~\cite{Guo:2024nzi}.
Additionally, the EOS of the NS crust is taken by interpolating to a BPS EoS~\cite{Baym:1971pw} below \(0.5n_0\)~\cite{Arnett:1977czg}.

Considering the well determined masses of mesons and nucleons from experiments~\cite{ParticleDataGroup:2024cfk}, we set \(m_{\omega}=782\ {\rm MeV}\), \(m_{\rho}=763\ {\rm MeV}\), \(m_{a_0}=980\ {\rm MeV}\), and \(m_N=939\ {\rm MeV}\) in the numerical process.
The mass of \(\sigma\) meson is less certain due to its broad width and strong coupling to pions, so we let it vary from \(400\) to \(800\ {\rm MeV}\) in our work.
And from the chiral nuclear force models~\cite{Ren:2016jna,Djukanovic:2006mc,Girlanda:2010ya,Polinder:2006zh}, the corresponding OBE couplings, \(g_{\sigma NN}\), \(g_{\omega NN}\), \(g_{\rho NN}\), and \(g_{a_0 NN}\) are limited within \(-20\) to \(20\).
From our previous works and the literatures~\cite{Ma:2025llw, Kovacs:2021ger, Parganlija:2012fy}, the coupling magnitudes of three-meson terms and four-meson terms are taken less than \(10000~\rm MeV\) and 1000, respectively.

Then, the model parameters \(\boldsymbol{\theta}\) to be discussed in the framework, from the Lagrangian in Ref.~\cite{Ma:2026kun}, are:
\( g_{\sigma N N}\), \( g_{\omega N N}\), \( g_{\rho N N}\), \( g_{a_0 N N}\), \( b_3\), \( b_4\), \( b_5\), \( b_6\), \( b_7\), \( b_8\), \( b_9\), \( c_3\), \( c_4\), \( g_2\), \( g_3\), \( g_4\), \( g_5\), \( g_6\), \( g_7\), \( d_3\), \( m_\sigma\).
Now, we can see have a 21-dimensional parameter space to be explored within the Bayesian analysis framework, and this will make the calculation of Eq.~\eqref{eq:evidence} computationally challenging, which leads to the discussion on how different prior distributions \(p(\boldsymbol{\theta})\) affect the analysis results and how model perform difficult.
So, in this work, we choose the uniform prior distributions for all model parameters within the physically reasonable ranges mentioned above, and use the Eq.~\eqref{eq:likelihood} in practice to be the benchmark for the model performance.

\section{Neutron star structure and sound velocity from the general Walecka-type model}
\begin{figure}[htbp]
    \centering
    \includegraphics[width=0.5\linewidth]{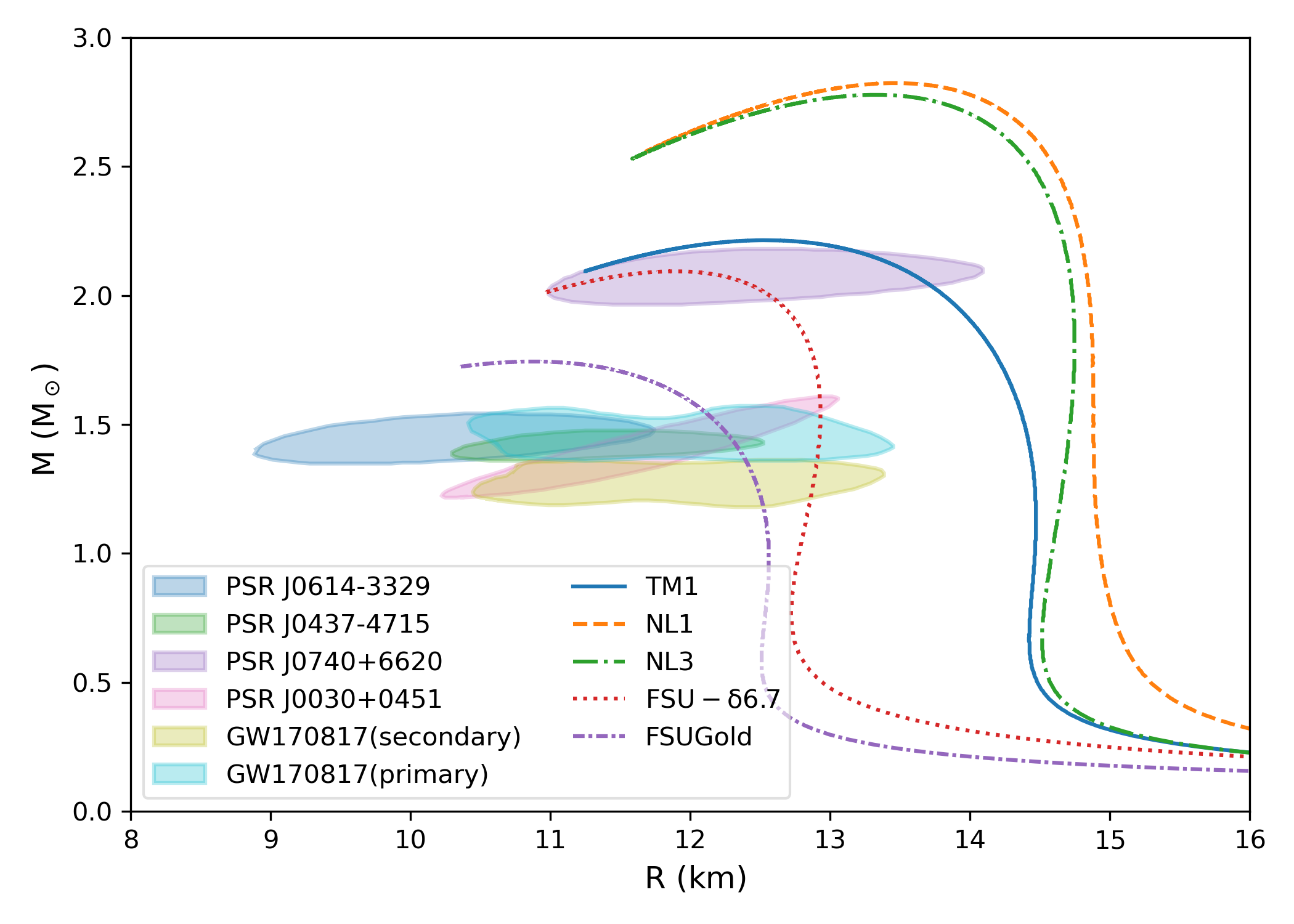}
    \includegraphics[width=0.5\linewidth]{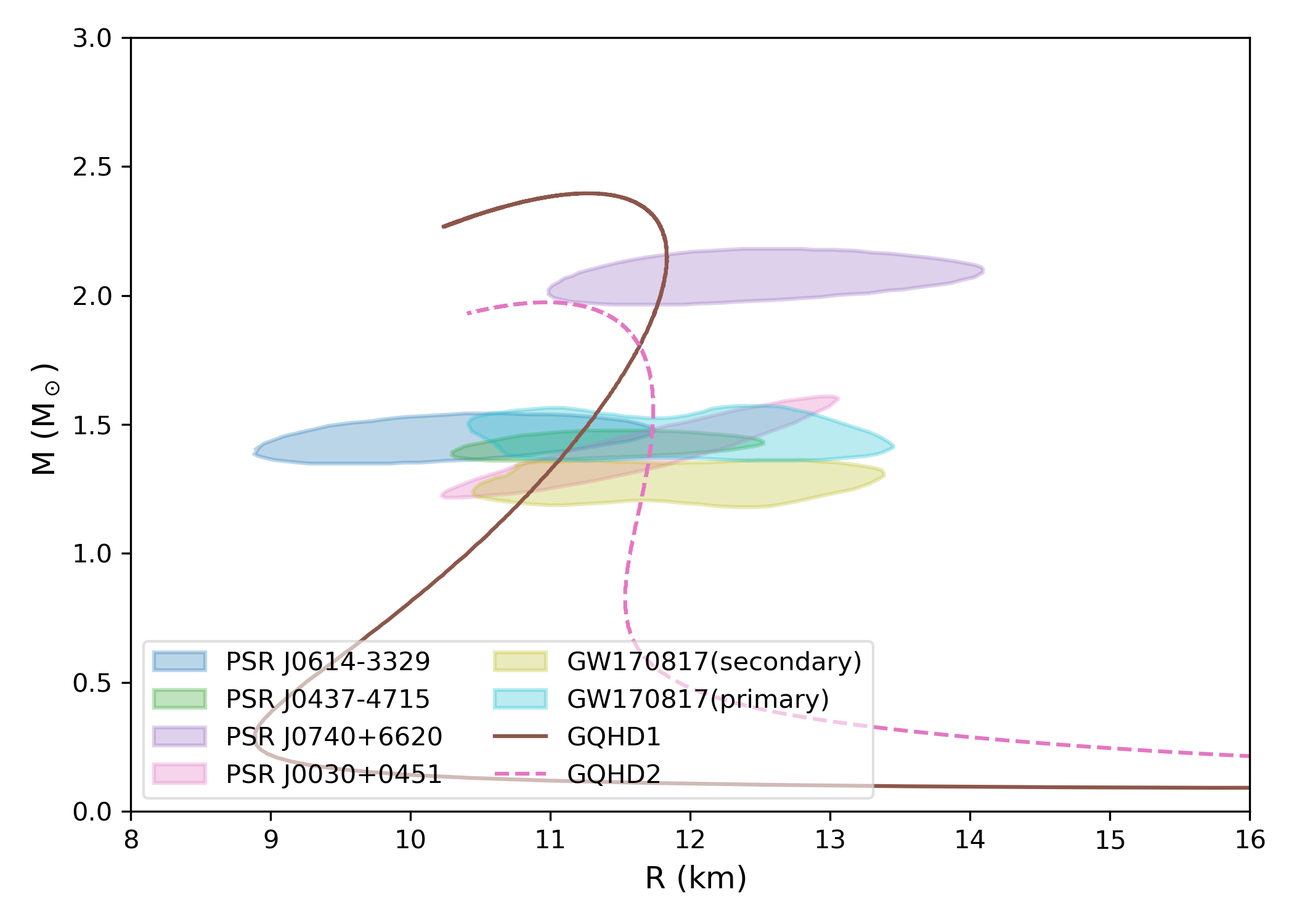}
    \caption{
        The M-R relation for different parameter sets with beta-equilibrium.
        The constraints PSR J1614-2230, PSR J0348+0432, PSR J0740+6620, J0030+0451, PSR J0437–4715, and PSR J0614–3329 are from Refs.~\cite{LIGOScientific:2017vwq,LIGOScientific:2017ync,LIGOScientific:2018cki, Salmi:2024aum, Vinciguerra:2023qxq, Choudhury:2024xbk, Mauviard:2025dmd}, respectively.
        }
    \label{fig:mr-relation}
\end{figure}

With the experimental and observational constraints considered, we found two optimal parameter sets GQHD1 and GQHD2 for the general Walecka-type model, whose details can be found in Ref.~\cite{Ma:2026kun}.
With well-reproduced nuclear matter properties, the resulting NS M-R relations, compared to past studies based on Walecka-type models~\cite{Sugahara:1993wz,reinhard1986nuclear,Lalazissis:1996rd,Li:2022okx,Todd-Rutel:2005yzo}, are shown in Fig.~\ref{fig:mr-relation}, respectively.
One can see that from Fig.~\ref{fig:mr-relation}, compared to past parameter sets, the GQHD1 and GQHD2 outperforms for NS M-R relation in terms of reproducing a more compact NS having mass of around \(1.4M_{\odot}\).

\subsection*{Sound velocity and the role of \(\sigma\omega\rho a_0\) term}

\begin{figure}[htbp]
    \centering
    \includegraphics[width=0.5\linewidth]{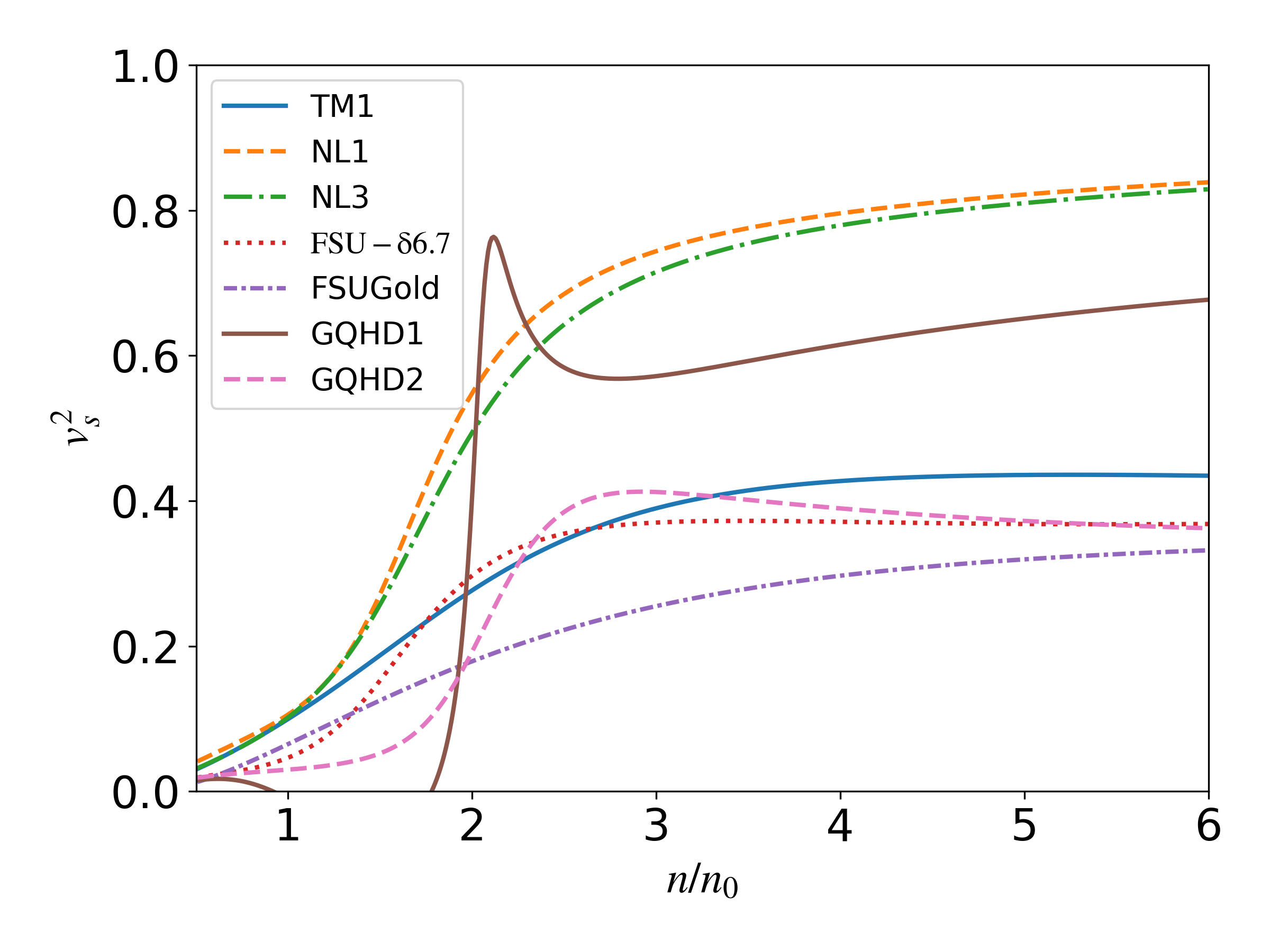}
    \caption{Sound velocity \(v_s^2\) vs. density for various Walecka parameter sets~\cite{Sugahara:1993wz,reinhard1986nuclear,Lalazissis:1996rd,Li:2022okx,Todd-Rutel:2005yzo,Ma:2026kun}.}
    \label{fig:speedOfSound}
\end{figure}

The speed of sound \(v_s^2\) provides information about EOS stiffness and important aspects of QCD~\cite{Tews:2018kmu,Bedaque:2014sqa,Altiparmak:2022bke,Fujimoto:2022ohj}. While peak structures were previously attributed to phase transitions, Fig.~\ref{fig:speedOfSound} shows that GQHD1 and GQHD2 exhibit peaks at intermediate densities within pure hadronic matter.

\begin{figure}[htbp]
    \centering
    \includegraphics[width=0.45\linewidth]{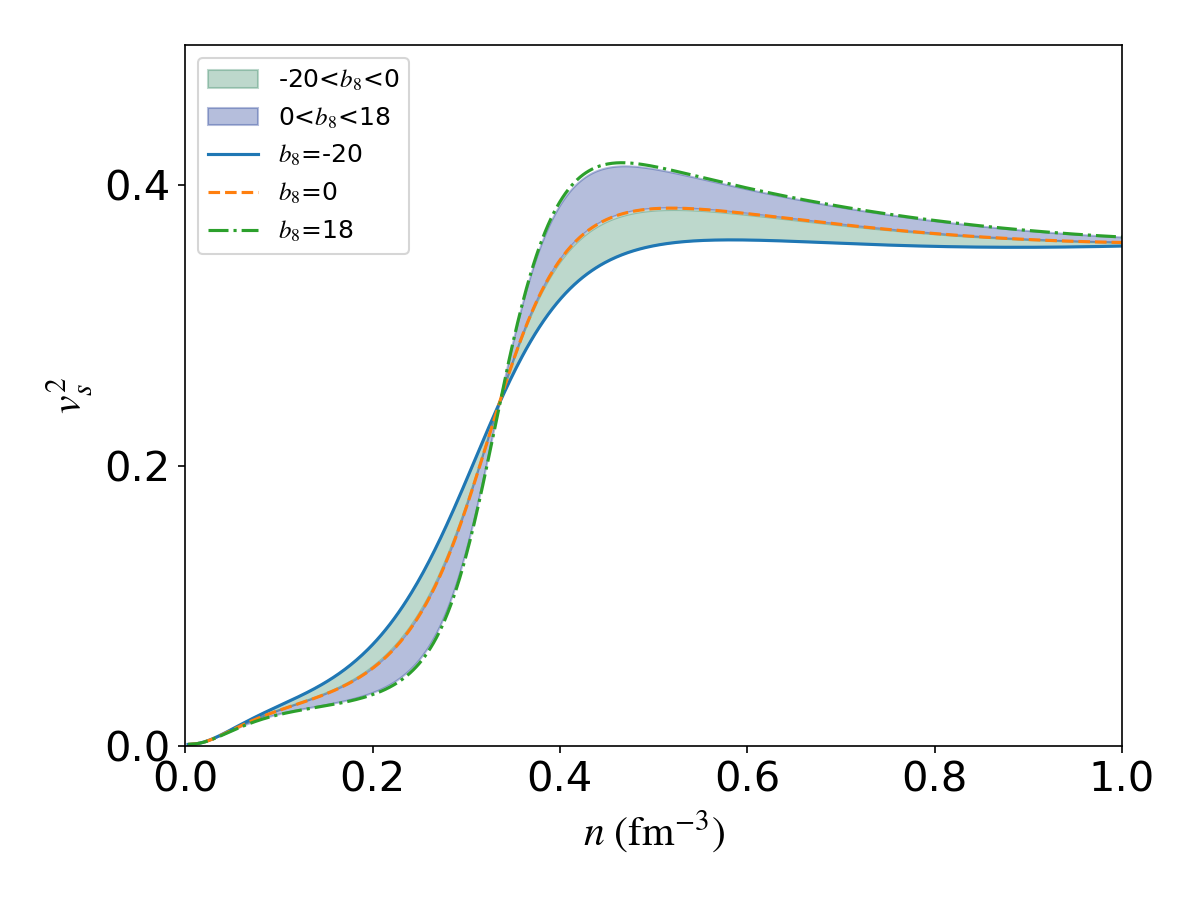}
    \includegraphics[width=0.45\linewidth]{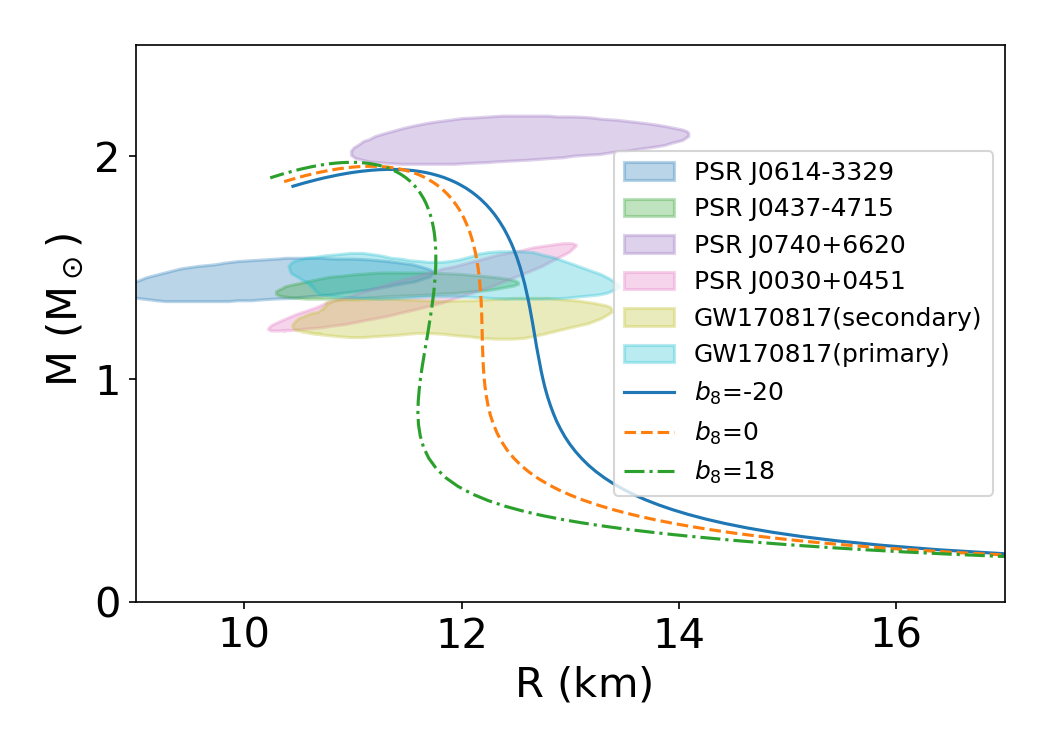}
    \caption{Sound velocity and M-R relation for different \(b_8\) values in GQHD2.}
    \label{fig:V12P8}
\end{figure}

We found that the mixing term \(b_8\sigma\omega\rho a_0\) is crucial for producing this peak. Figure~\ref{fig:V12P8} demonstrates that as \(b_8\) increases, the peak emerges and reduces NS radii around \(1.4M_{\odot}\) while preserving the \(2M_{\odot}\) constraint.
Besides, at high densities, \(v_s^2\) approaches the conformal limit \(1/3\).

\section{Summary and outlook}

In this work, we have developed a general Walecka-type RMF model for dense nuclear matter and performed a comprehensive Bayesian analysis to explore its parameter space under constraints from nuclear matter properties and neutron star observations.
And the results indicate that the general Walecka model includes some operators that can effectively capture the high density behaviors but not considered by past studies.
With the general Walecka model and an appropriate choice of parameters, it is possible to make a NS with mass of around \(1.4M_{\odot}\) have a rather small radius to satisfy the possible constraints from PSR J0614–3329.
And results of sound velocity not only demonstrate the role of the sound speed peak in determining the NS structures, but also highlights the importance of the mixed interaction term \(b_8\sigma \omega \rho a_0\) in shaping EOS at intermediate densities.
This also provides a new perspective on the microscopic origin of the sound speed peak just with pure hadronic matter.

In the future, we plan to further extend our analysis framework to studies on NSs with chiral effective field theories established in Refs.~\cite{Ma:2025llw,Zhang:2024sju}, which can provide a more clear connection to the underlying QCD dynamics, to further explore the microscopic origin of the sound speed peak and its implications for NS structures.

\section*{Acknowledgements}
The work of Y. M. is supported by by Jiangsu Funding Program for Excellent Postdoctoral Talent under Grant Number 2025ZB516.

The authors would like to thank the useful discussions with Dr. Ling-Jun Guo. 

\bibliographystyle{IEEEtran}
\bibliography{AI}
\end{document}